\documentclass[preprint,showpacs,superscriptaddress,nofootinbib,preprintnumbers,11pt]{revtex4-1}

\usepackage{array}
\usepackage{amsmath}
\usepackage{hyperref}
\usepackage{mathrsfs}
\usepackage{breakurl}
\usepackage{amsmath}
\usepackage{graphicx,epsfig}
\usepackage{bm}
\usepackage{latexsym,amssymb,amsmath}
\usepackage{url}
\newcommand{\beq}[1]{\begin{equation}\label{#1}}
\newcommand{\eeq}{\end{equation}}
\newcommand{\bea}{\begin{eqnarray}}
\newcommand{\eea}{\end{eqnarray}}
\newcommand{\ba}{\begin{array}}
\newcommand{\ea}{\end{array}}

\newcommand{\rf}[1]{(\ref{#1})}
\def\be{\begin{equation}}
\def\ee{\end{equation}}
\def\gs{\mathrel{
   \rlap{\raise 0.511ex \hbox{$>$}}{\lower 0.511ex \hbox{$\sim$}}}}
\def\ls{\mathrel{
   \rlap{\raise 0.511ex \hbox{$<$}}{\lower 0.511ex \hbox{$\sim$}}}}

\newcommand{\lsim}{\mathrel{\vcenter{\hbox{$<$}\nointerlineskip\hbox{$\sim$}}}}
\newcommand{\gsim}{\mathrel{\vcenter{\hbox{$>$}\nointerlineskip\hbox{$\sim$}}}}

\newcommand{\half}{\frac{1}{2}}
\newcommand{\third}{\frac{1}{3}}
\newcommand{\quarter}{\frac{1}{4}}

\newcommand{\rarr}{\rightarrow}

\def\ie{i.e.\ }

\newcommand{\bad}{\begin{array}{ccc}}
\newcommand{\bav}{\begin{array}{cccc}}
\newcommand{\baf}{\begin{array}{ccccc}}

\newcommand{\nubar}{\bar\nu}
\newcommand{\nue}{\nu_{e}}
\newcommand{\numu}{\nu_{\mu}}
\newcommand{\nutau}{\nu_{\tau}}

\newcommand{\mutau}{\nu_\mu$-$\nu_\tau}

\newcommand{\wtr}{w_{\rm track} }
\newcommand{\wsh}{w_{\rm shower} }
\newcommand{\Ntr}{N_{\rm track} }
\newcommand{\Nsh}{N_{\rm shower} }
\newcommand{\Ntot}{N_{\rm Total} }

\newcommand{\al}{\alpha}

\newcommand{\scrP}{\mathscr{P}}

\begin{document}

\preprint{NSF-KITP-15-002}

\title{Aspects of the Flavor Triangle for Cosmic Neutrino Propagation}

\author{Lingjun Fu}
\email[Email: ]{lingjun.fu@vanderbilt.edu}
\affiliation{Department of Physics and Astronomy, Vanderbilt University, Nashville, TN 37235, USA}
\author{Chiu Man Ho}
\email[Email: ]{cmho@msu.edu}
\affiliation{Department of Physics and Astronomy, Michigan State University, East Lansing, MI 48824, USA}
\author{Thomas J. Weiler}
\email[Email: ]{tom.weiler@vanderbilt.edu}
\affiliation{Department of Physics and Astronomy, Vanderbilt University, Nashville, TN 37235, USA}

\date\today

\begin{abstract}
\noindent
Over cosmic distances, astrophysical neutrino oscillations average out to a classical
flavor propagation matrix $\scrP$. Thus, flavor ratios injected at the cosmic source $W_e,W_\mu,W_\tau$ evolve to
flavor ratios at Earthly detectors $w_e,w_\mu,w_\tau$ according to $\vec{w}=\scrP \vec{W}$.
The unitary constraint reduces the Euclidean octant to a ``flavor triangle''.
We prove a theorem that the area of the Earthly flavor triangle is proportional to Det$(\scrP)$.
One more constraint would further reduce the dimensionality of the flavor triangle at Earth (two) to a line (one).
We discuss four motivated such constraints.
The first is the possibility of a vanishing determinant for $\scrP$.
We give a formula for a unique $\delta(\theta_{ij}$'s) that yields the vanishing determinant.
Next we consider the thinness of the Earthly flavor triangle.
We relate this thinness to the small deviations of the two angles $\theta_{32}$ and $\theta_{13}$ from
maximal mixing and zero, respectively.
Then we consider the confusion resulting from the tau neutrino decay topologies, which are showers at low energy,
``double-bang'' showers in the PeV range, and a mixture of showers and tracks at even higher energies.
We examine the simple low-energy regime, where there are just two topologies, $\wsh=w_e+w_\tau$ and $\wtr=w_\mu$.
We apply the statistical uncertainty to be expected from IceCube to this model.
Finally, we consider ramifications of the expected lack of $\nutau$~injection at cosmic sources.
In particular, this constraint reduces the Earthly triangle to a boundary line of the triangle.
Some tests of this ``no~$\nutau$~injection'' hypothesis are given.
\end{abstract}

\pacs{14.60.Pq, 95.55.Vj, 95.85.Ry}

\maketitle

\section{Introduction to Cosmic Neutrino Flavors and Flavor Triangles}
\label{sec:intro}
The IceCube neutrino telescope has begun observation of neutrinos from distant sources~\cite{Aartsen:2014gkd,*Aartsen:2013jdh}.
It is expected that detections from the recently deployed Antares~\cite{Antares}, and the soon-to-be deployed 
KM3NeT~\cite{KM3NeT}, will soon follow.
After an ensemble of neutrino events have been collected, track versus shower topologies will allow
one to extract a neutrino flavor ratio arriving at Earth~\cite{Beacom:2003nh}. 
In the past three years, IceCube has announced three showering events characteristic of $\nue$'s or $\nutau$'s
(or their antiparticles, since non-magnetized neutrino telescopes cannot distinguish $\nu$ from $\nubar$), 
in the energy range $\sim$1--2~PeV, in addition to another 34 events with energies between 30 TeV and 300 TeV. 
At 5.7$\sigma$, a purely atmospheric neutrino background explanation of these events is rejected.
We have likely witnessed the first observations of high-energy extra-galactic neutrinos. 

In the standard treatment of neutrino oscillations, the neutrino states in the flavor and mass bases
are related by a unitary transformation $|\nu_\al \rangle =\sum_{j}\, U_{\al j}^\ast\, |\nu_j \rangle $.\,
where $\al=e,\,\mu,\,\tau$ and $j=1,\,2,\,3$ are the indices for the flavor and mass eigenstates, respectively. The unitary transformation
is described by the Pontecorvo-Maki-Nakagawa-Sakata (PMNS) matrix $U$ whose elements are given by $U_{\al j}=\langle \nu_\al | \nu_j\rangle$.
For a given propagation distance $L$, the flavor state $|\nu_\al \rangle $ evolves into
$|\nu_\al (L) \rangle =\sum_{k}\, e^{-i\,E_k\,L}\,U_{\al k}^\ast\, |\nu_k \rangle$.
The transition probability from the flavor state $|\nu_\al (L)\rangle$ to $|\nu_\beta \rangle$ is then given by
$P_{\al\beta}= |\langle \nu_\beta|\nu_\al(L) \rangle|^2$ for any $\alpha$ and $\beta$ taken from the set $\{e,\,\mu,\,\tau\}$.

For cosmic neutrinos, the characteristic distance is much larger than the oscillation length, 
motivating a statistical average over a neutrino ensemble. 
This averaging randomizes a quantum-mechanical phase $\phi_{jk}\,\equiv \,L\,(m_j^2-m_k^2)/2E$ between states
such that the phase factor $e^{i\phi_{jk}}$ averages to zero (think of wrapping a complex number of unit modulus 
around zero in its complex plane).
The result is a reduction of $P$ to $\langle P \rangle \equiv \scrP$, 
where the brackets connote the averaging of all phase factors to zero.
The matrix elements of the reduced propagation matrix $\scrP$ are simple, positive definite elements:
\beq{Pab}
\scrP_{\alpha\beta}= \sum_j \, | U_{\alpha j} |^2\, | U_{\beta j} |^2 = \left(\,|U|^2\, \left(\,|U|^2\,\right)^T\,\right)_{\alpha\beta}\,,
\eeq
%
%
and the matrix elements of $|U|^2$ are defined to be $|U_{\alpha\beta}|^2$.
Expressions for the elements $\scrP_{\alpha\beta}$ in terms of the PDG set 
\{$\theta_{32},\ \theta_{12},\ \theta_{13},\ \delta$\}, are given in Ref.~\cite{pakvasa:2007dc}.
It is this $\langle P\rangle =\scrP$~matrix that propagates the flavor ratios injected at the source, 
$\,\vec{W}=(W_e,W_\mu,W_\tau)$ with normalization $W_e+W_\mu+W_\tau=1$,
to the flavor ratios observed on Earth, 
$\,\vec{w}=(w_e,w_\mu,w_\tau)$ with mormalization $w_e+w_\mu+w_\tau =1$,
\ie $\vec{w}=\scrP \,\vec{W}$.

\parskip=0pt 
We can understand the formula in Eq.~\rf{Pab} 
as follows. 
The correct basis for neutrino propagation is the mass
basis, as dictated by the particle propagator in field theory, an analytic function with poles at mass values.
The mass states labeled by $j$ are unobserved and so we need to sum over them. 
Furthermore, each of these propagating mass states is weighted by its classical probability $|U_{\alpha j}|^2$ 
to overlap with the flavor neutrino $|\nu_\alpha \rangle$ produced at the source,
times its classical probability $|U_{\beta j}|^2$ to overlap with flavor neutrino $|\nu_\beta \rangle$ detected at Earth. Since phase-averaging restores $CP$-invariance, the matrix $\scrP$ describes both neutrino and anti-neutrino propagation equally.
Furthermore, according to the $CPT$-theorem, $CP$-invariance also implies $T$-invariance,
and so the matrix $\scrP$ is symmetric, namely $\scrP_{\alpha\beta} = \scrP_{\beta\alpha}$.

\parskip=0pt 
The outline for this paper is:
In \S\rf{sec:perturbedTBM} we discuss the TriBiMaximal (TBM) matrix, 
and additional perturbations that are needed to agree with experiment.
We argue that the TBM matrix offers a $\scrP$ matrix that is qualitatively good, 
but not good in the details warranted by some observables.  
One feature of $\scrP_{\rm TBM}$ that 
will turn out to be significant is the vanishing determinant.
With a vanishing determinant, one cannot invert the propagation to infer the cosmic flavor ratios from the measured, Earthly ones.
We present the restrictions on the flavor propagation matrix $\scrP$ that result 
from unitarity of the PMNS leptonic mixing matrix.
In \S\rf{sec:TriEarth} we present the concept of flavor triangles at Earth
as a concise way to indicate the result of flavor decoherence arising from propagation of neutrinos over cosmic distances.
Some illustrative examples of $\scrP$ matrix and their associated Earthly triangles are given. In \S\rf{sec:line} we expand on the idea that just as the constraint $w_e+w_\mu+w_\tau=1$ reduces a Euclidian volume octant to 
a triangular surface, so will another constraint reduces the triangle to a line.  We give four physically motivated examples 
of such constraints, the first two involving properties of the propagation matrix (equivalently, constraints from or on the mixing angles), 
the third involving Earthly detector efficiencies, and the fourth involving particle physics at the sources.
Our first example, that of a vanishing determinant of $\scrP$, is presented in  \S\rf{subsec:VanishingDet}.
The TBM ans\"atz gives a vanishing determinant.  Since TBM is known phenomenologically to 
be nearly true, the true determinant cannot be far from zero.  We show that it can be zero, 
and that the breaking of the $\mutau$~symmetry does not necessarily imply a non-vanishing Det($\scrP$).
We discuss implications for $\numu$-$\nutau$ symmetry, the cornerstone of the TBM ans\"atz.  
We derive the unique value of $\delta$ that simultaneously allows for vanishing Det($\scrP$) and broken $\mutau$~symmetry.
Finally, we prove a theorem that the area of the Earthly flavor triangle and the determinant of $\scrP$
are directly proportional, with proportionality constant $\sqrt{3}/2$.
This theorem therefore tells us that any nonzero area for the Earthly triangle implies an invertible $\scrP$,
which in turn implies the possibility of reconstructing the flavor ratios at injection directly from the ratios measured at Earth.
In \S\rf{subsec:thinness} we present our second example of ``constraint'' that evokes a straight line solution for the Earthly flavors,
namely, the conditions for a triangle that is so thin that it is experimentally indistinguishable from a straight-line solution. It turns out that the width of the triangle projected onto the $w_\mu$-$w_\tau$-axis, 
which is zero with good $\mutau$~symmetry,
is second order in the standard angle-phase parameter $\theta_{13}\,\cos\delta$
and second order in the deviation of the $\theta_{32}$ from its TBM value of $\pi/4$. 
Since experiment tells us that both of these parameters are small, their squares are very small, 
and we understand the thinness of the Earthly triangle.
In \S\rf{subsec:2wandStatErrors} we discuss the 
ambiguities presented by the $\nutau$ events.
At energies $\lsim$~PeV, the $\nutau$~events look like showers.
At energies $\gsim 10$~PeV, the double-bang nature of the $\nutau$ events leads to a combination of shower, track 
and identifiably $\nutau$ events. We consider the simplified case for $E_\nu\lsim$~PeV, 
where only two event topologies of ``flavor'' are resolved in Earthly detectors, 
namely ``tracks'' and ``showers''.  
This simplification then becomes our third example of a constraint which reduces the flavor triangle to a line.
We assess the statistical significance of the flavor inferences of this model, 
for the case of 140 total events and an injection ratio of (1:2:0).  
The 140 event sample is the typical expectation for 10~years of IceCube running,
or 1-2 years of running for IceCube's proposed, large-volume extension, Gen2.  
The statistical limitations on the determination of $\vec{w}$ are significant.
In \S\rf{subsec:nonutau} we discuss the constraint that results from assuming that the injection of $\nutau$'s is negligibly small, 
as is expected in all popular models.
Although the Earthly triangle may have a nonzero area from considerations of the $\scrP$~matrix, 
the ``no~$\nutau$~injection'' constraint reduces the Earthly possibility to just a boundary line of the Earthly triangle.
Differentiation of this boundary line from the thin flavor triangle requires a considerable set of statistics (events).
Such an event collection may not be achievable in the near future.
Our conclusions are collected in \S\rf{sec:conclusions}.

\section{Perturbations About TBM Values, and Flavor Triangles}
\label{sec:perturbedTBM}
With the TBM model, the $U$ matrix, the $|U|^2$ and $\scrP$ matrices, are
\beq{UsquaredTBM}
U_{\textrm{TBM}}=\frac{1}{\sqrt{6}}
\left(
\ba{ccc}
    2  & \sqrt{2} &        0         \\
 -1 & \sqrt{2} &  \sqrt{3}  \\
 -1 & \sqrt{2} & -\sqrt{3}
\ea
\right)\,,\quad
|U|^2_{\textrm{TBM}} = \frac{1}{6}
\left(
\ba{ccc}
 4 & 2 & 0 \\
 1 & 2 & 3 \\
 1 & 2 & 3
\ea
\right)\,,\quad{\rm and}\ \ 
%
\scrP_{\rm TBM}=\frac{1}{18}
\left(
\ba{ccc}
10 & 4 & 4  \\
 4  & 7 & 7 \\
 4  & 7 & 7 \\
\ea
\right)\,.
\eeq

Two ingredients of the TBM ans\"atz that are necessary to realize the $\numu$-$\nutau$ symmetry are that $\theta_{13} = 0$, forcing $U_{e3}$ to be zero, and that $\theta_{32}$ is maximal, equal to $\frac{\pi}{4}$.
However, the DAYA-BAY experiment in China \cite{An:2012eh} has inferred a nonzero 
$\sin^2\,(2 \theta_{13}) = 0.092 \pm 0.016\, (\textrm{stat.}) \pm 0.005\, (\textrm{syst.})$, 
and and the RENO experiment in Korea \cite{Ahn:2012nd} has 
$\sin^2\,(2 \theta_{13}) = 0.113 \pm 0.013\, (\textrm{stat.}) \pm 0.019 \,(\textrm{syst.})$, each at 68\% C.L. 
These values of $\sin^2(2\theta_{13})$ give a $\theta_{13}$ that is more than $10\sigma$ removed from zero, 
indicating that the TBM model is not valid. Indeed, cosmic neutrinos can be used as a probe of broken $\nu_\mu$-$\nu_\tau$ 
symmetry~\cite{Xing:2006xd}. 

In 3D space, the allowed area of the flavor fractions at injection is an equilateral hyper-triangle, with vertices at $(1,0,0)$, $(0,1,0)$, and $(0,0,1)$ 
in the Euclidean $W_e$--$W_\mu$--$W_\tau$ space. 
The allowed Earthly fractions, with unitary constraint $w_e+w_\mu+w_\tau=1$, 
constitute a flavor triangle defined by the three vertices at
$\scrP\,(1,0,0)^{\rm T}$, $\scrP\,(0,1,0)^{\rm T}$, and $\scrP\,(0,0,1)^{\rm T}$. 
The unitary and symmetric properties of $\scrP$ matrix are encapsulated in writing:
\bea
\label{P-sym}
\scrP=
\left(
\ba{ccc}
1-(a+b) & a & b  \\
a & 1-(a+c) & c \\
b & c & 1-(b+c) \\
\ea
\right)
=\frac 1 {18} \left(
\ba{ccc}
10 & 4 & 4  \\
4 & 7 & 7 \\
4 & 7 & 7 \\
\ea
\right)
+\frac{1}{18}\,\Delta {\scrP}\,,
\eea
where the upper bound on the off-diagonal, flavor-changing probabilities $\scrP_{\alpha\beta}$ is $\half$ 
from the two-flavor oscillation limit,
and the well-known lower bound on the diagonal matrix elements $\scrP_{\alpha\alpha}$ for the three-flavor system is $\third$;
these limits lead directly to $0\le a,\ b,\ c \le \half$,
and the pairwise range $0\le a+b,\ b+c,\ c+a \le \frac{2}{3}$.\footnote
{We note that unitarity relations were recently presented in detail in~\cite{Xu:2014via}. 
Our simple constraints here are equivalent to two of the three unitarity results obtained there.
Their third constraint, that twice any element of the set $\{a,b,c\}$ 
plus either one of the remaining two elements is bounded by $\frac{25}{24}$
(for example, that $2a+b\le\frac{25}{24}$), is neither reproduced nor needed here.
}
 
Here, $\Delta \scrP$ is the perturbation over the TBM matrix.
The expansion of $\Delta\scrP$ to second order in the deviations of the three leptonic mixing angles
from their TBM values can be found in~\cite{Pakvasa:2008zz}. There it was derived that
\beq{deltaP}
\Delta\scrP = A\,
\left( 
\ba{ccc}
4 & -2 & -2  \\
-2 & 1 & 1 \\
-2 & 1 & 1 \\
\ea
\right)
+B\,
\left( 
\ba{ccc}
0 & 1 & -1  \\
1 & -1 & 0 \\
-1 & 0 & 1 \\
\ea\right)
+C\,
\left(
\ba{ccc}
0 & 0 & 0  \\
0 & 1 & -1 \\
0 & -1 & 1 \\
\ea\right) \,,
\eeq
where $A,\,B$ and $C$ are calculated functions of the deviations of mixing parameters from the TBM values.
$A$ does not depend on $\delta$, and in the best-fit case~\cite{Forero:2014bxa} is equal to -0.154;
$B$ ranges over $[-0.003,\,-0.842]$ as $\cos\delta$ ranges over $[-1,\,1]$;
$C$ is quadratic in the deviation, to lowest order ($A$ and $B$ each contain a linear dependence), and ranges over $[0.073,\,0.319]$.
 We note that the perturbations to the TBM matrix are much smaller than the elements of the TBM matrix:
 The $A$-correction is at most 8\%, the $B$-correction is at most 20\%, and the $C$-correction is at most 5\%.
Thus, the TBM ordering $\scrP_{ee} > \{\scrP_{\mu\mu},\ \scrP_{\mu\tau},\ \scrP_{\tau\tau} \} > \{\scrP_{e\mu},\ \scrP_{e\tau} \}$
(i.e., $1 > a+b+c$, and $c > a {\rm\  or\ } b$)
is maintained in the real world.  
This ordering result will be an important ingredient in establishing a conclusion in \S\rf{subsubsec:smalltau}.
We have shown that the TBM matrix is a good approximation, and in much of what follows we adopt this approximation.
However, it has a vanishing determinant, and so cannot serve as an approximation for discussions that require 
a non-vanishing determinant.
 
\section{The Flavor Triangle at Earth}
\label{sec:TriEarth}
It is clear that nonzero $a,\ b,\ c$, i.e.\ neutrino flavor mixing, reduce the size of the Earthly triangle relative to the original injection triangle.

\subsection{The Centroid Point}
\label{susec:centroid}
The centroid of the Earthly triangle is at the point ($\third,\ \third,\ \third$), as shown in Fig.~\ref{fig:quark-mixing}. 
It can be achieved by symmetric mixing, $a=b=c=\third$, which further implies the democratic matrix
with all elements equal to $\third$:
\beq{third-symmetry}
\scrP_{\third}=\third
\left(
\ba{ccc}
1 & 1 & 1 \\
1 & 1 & 1 \\
1 & 1 & 1 
\ea
\right)\,.
\eeq
Unitarity guarantees that any injection vector $\vec{W}$ will return the centroid point at Earth when propagated by $\scrP_{\third}$.
The Determinant of $\scrP_{\third}$ vanishes, so $\scrP_{\third}$ is not invertible;
while all input vectors $\vec{W}$ will lead to an Earthly (1:1:1) ratio were this propagation matrix to be correct, 
the converse is not true -- an Earthly (1:1:1) ratio does not imply the correctness of this propagation matrix.
A common counter-example is the input vector resulting from charged-pion decays that $\vec{W}=\third$(1:2:0),
which leads to $\third$(1:1:1) if a propagation matrix close to the TBM matrix, given in Eq.~\rf{UsquaredTBM},  is assumed.

We find that all Earthly triangles determined by arbitrary 
$\theta_{32},\ \theta_{21}$, and the pair ($\theta_{13},\ \delta_{\rm CP}$), include the centroid point. 
The reason is simple: the unitarity constraint guarantees that the big injection triangle and small Earthly triangle share the same centroid. 
Of course, restrictions on the initial values of the $\vec{W}$ flavor components may restrict the Earthly triangle 
to an area that does not contain the centroid. 
An example of such a restriction is the common assumption that little or no $\nutau$'s are produced at the source,
i.e., that $W_\tau$ is effectively zero.  We discuss the implications of this assumption in a later section.

\subsection{Example: Quark Mixing}
\label{subsec:Qmix}
\begin{figure}[t!]
\includegraphics[height=6.5cm, width=7.5cm]{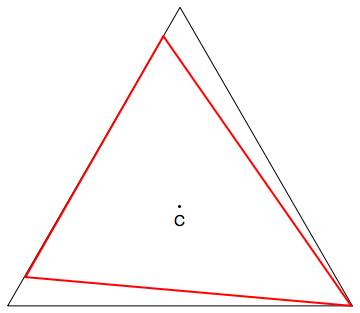}
~~~~~~
\caption{The analog of the Earthly triangle is shown (red interior triangle) for the mixing angles that relate quark flavors and masses.
Also shown is the centroid point, labelled ``C''.}
\label{fig:quark-mixing}
\end{figure}

It is well known that the quark mixing angles are much smaller than those of the  neutrino sector.
Hence, we expect the analog of quark flavor mixing to offer an Earthly triangle much more similar to the unmixed source triangle.
In Fig.~\ref{fig:quark-mixing}, we visualize the smallness of the quark mixing angles by plotting the 
reduction of the hadronic source triangle to
what would be the Earthly triangle if quarks were to oscillate 
(they don't, as they are confined, and their mass differences are very large).
The corners of the Earthly matrix are given by $\scrP$-propagation of the unmixed flavor vectors (1,0,0), (0,1,0), and (0,0,1).
As we discuss later, the quark triangle after mixing is about 80\% in area of the original unmixed triangle.

\section{One More Constraint Reduces the Triangle to a Line --\mbox{ Four Examples}}
\label{sec:line}
Since the unitarity constraint $\omega_e+\omega_\mu+\omega_\tau=1$ reduces a 3D volume to a 2D hyper surface
(in this case, to our flavor triangle), it is not surprising that a further constraint would reduce the 2D triangle to a 1D line.
We can think of four possible, interesting constraints, each of which would affect the dimensional reduction.
The first is a {\bf vanishing determinant of \boldmath{$\scrP$}}.
The second is that {\bf the triangle is so ``thin''} that no experiment will be able to differentiate the thin triangle from a straight line,
say, the line determining either of the thin triangle's longer borders.
The third is that in first approximation, a neutrino telescope can {\bf distinguish only between track and shower events} at neutrino energies 
$\lsim 1$~PeV or $\gsim 10$~PeV, and not among all three neutrino flavors.
And the fourth is the probably true statement that the sources do not emit any significant amount of $\nutau$'s, \ie {\bf ``no-$\nutau$~injected''}.
Of course, if two of these conditions hold, then the two lines will in general intersect in a point within the triangle,
and the flavor ratios at Earth will be determined.
In general, three of these conditions cannot hold simultaneously, unless they are linearly dependent.

The first two conditions, if true, result from properties of the $\scrP$ matrix.
The third condition results from a property of the detectors, their efficiency to identify ``double-bang'' $\nutau$ events~\cite{Learned:1994wg}.
And the final, fourth condition results from a property of the source, the presence or absence of a significant nonzero $W_\tau$.

There may also be new neutrino physics~\cite{Xing:2008fg,*Fu:2014gja}, such as mixing with sterile neutrinos, of new neutrino interactions,
but we consider these possibilities to be less motivated and less measurable in the flavor ratios.

In the following sections, we consider, one at a time, all four possible conditions. 
Since the Earthly triangle, once let loose subject to only unitarity, must contain the centroid point,
the first two conditions, studied in sections~\rf{subsec:VanishingDet} and ~\rf{subsec:thinness},  
will create a (one-dimensional) line through the centroid, 
but rotated in the plane relative to the vertical TBM line.
This rotation angle may be thought of as a measure of $\mutau$~symmetry breaking.
The third condition, studied in section~\rf{subsec:2wandStatErrors}, will reduce the Earthly triangle to a boundary line of the triangle, the line connecting 
the vectors $\scrP\,(1,0,0)^{\rm T}$ and $\scrP\,(0,1,0)^{\rm T}$.
This line is again rotated with respect to the vertical TBM line, but this time without passing through the centroid point.
The fourth possible condition, described and analyzed in section~\rf{subsec:nonutau}, reduces Earthly measurements to just two even topologies,
$\wtr$ and $\wsh$.

\subsection{Extra Constraint -- First Example: A Vanishing Determinant}
\label{subsec:VanishingDet}
An important, possible use of the $\scrP$~matrix is to use it to evolve backwards
the observed neutrino flavor ratio at Earth to obtain the ratios injected at the sources. 
The injection ratios then reveal the nature of source dynamics~\cite{Barger:2014iua,*Hummer:2010ai}. 

For a given neutrino flavor vector ${\vec W} \equiv (W_e, W_\mu, W_\tau)$ injected at cosmic sources,
the corresponding flavor vector ${\vec w}\equiv (w_e, w_\mu, w_\tau)$ measured at Earth is given by ${\vec w} = \scrP\,{\vec W}$.
%
%
If $\scrP$ has a non-vanishing determinant and hence is an invertible matrix, then the
inverse equation
\beq{FlavorProp2}
{\vec W} = \scrP^{-1}\,{\vec w}
\eeq
allows one to use the neutrino flavor ratios observed at Earth to determine those dynamically injected at cosmic sources~\cite{Fu:2012zr}. 
From Eq.~\rf{Pab} we get that the determinant of the propagation matrix $\scrP$ is given by
\bea
\label{DetP}
{\rm Det}(\scrP)={\rm Det}\left(\,|U|^2 \boldmath \cdot (|U|^2)^T\,\right)=\left(\,{\rm Det}\,(|U|^2)\,\right)^2\,.
\eea
Inversion of the $\scrP$ matrix was not viable with the TBM ans\"atz for neutrino mixing, 
because the nature of the $\mutau$~symmetry assumed for the matrix was expressed as identical second and third rows 
up to phases ($\pm 1$)~\cite{Harrison:2002et, *Harrison:2002kp}.  
This implies identical second and third rows for the $|U|^2$~matrix, and therefore a vanishing determinant.  
Since the determinant of $\scrP$ is equal to the squared determinant of $|U|^2$ (as given in Eq.~\rf{DetP}),
%
%
with the TBM ans\"atz, $\scrP$ has a vanishing determinant and is therefore not invertible.

Backwards evolution, as spelled out in Eq.~\rf{FlavorProp2}, requires that the matrix $\scrP$ be invertible,
i.e., have a nonzero determinant.
However, if the $\numu$-$\nutau$ symmetry were exact, the determinant of the $\scrP$~matrix would vanish.
We now know that $\theta_{13}$ is nonzero.
A common belief is that nonzero $\theta_{13}$ implies that the $\numu$-$\nutau$ symmetry is broken.
However, utilizing the PDG form of the PMNS lepton-mixing matrix, we may conclude that 
$\numu$-$\nutau$ symmetry arises from the following conditions~\cite{Liao:2012xm}: 
(i) $\theta_{32}=\frac \pi 4$, as in the TBM ans\"atz, and 
(ii) $\sin 2\theta_{21}\sin \theta_{13}\cos \delta=0$. 
Given that inferences from neutrino oscillation experiments are that $\theta_{13}\neq 0$ and $\theta_{21} \neq \frac \pi 2$, 
the only remaining possibility for exact $\numu$-$\nutau$ symmetry is that 
$\theta_{32}=\frac \pi 4$, and $\delta = \pm\,\pi/2$. 
Although the restriction of $\delta$ to $\pm\frac{\pi}{2}$ is not part of the TBM ans\"atz,
this particular value $\delta=\pm\frac{\pi}{2}$ is presently viable.\footnote{In a recent data fit~\cite{Gonzalez-Garcia:2014bfa}, 
slightly favored values of $\delta$ are $\pm\,\pi/2$.  
However, all values of $\delta$ are presently permissible at $2\sigma$ range.}
Nevertheless, it seems probable at present that $\mutau$~symmetry is broken.

Even so, a point which we choose to emphasize is that broken $\mutau$~symmetry
does not imply that the determinant for $\scrP$ is non-vanishing.
One of the major points we explore in this paper is the possibility of having a vanishing determinant for $\scrP$ 
even if the $\mutau$~symmetry is broken.

\subsubsection{Det($\scrP$), Vanishing or Not? }
\label{subsec:DetP}
%
%
According to Eq.~\rf{DetP}, to study the vanishing of ${\rm Det}(\scrP)$, 
it is enough to analyze the vanishing of ${\rm Det}(|U|^2)$. 
In linear algebra, ${\rm Det}(|U|^2)=0$ means that the three rows of $|U|^2$ 
(or columns, since $|U|^2\rarr ( |U|^2)^{\rm T}$ leaves the determinant invariant) are linearly-dependent~\cite{LinearAlgebra}. Explicitly, one has the three equations
\beq{constant}
\alpha\,|U_{e j}|^2+\beta\,|U_{\mu j}|^2+\gamma\,|U_{\tau j}|^2=0\quad{\rm for\ }j=1,\ 2,\ 3\,,
\eeq
where at least two of the three constants $\alpha,\beta,\gamma$ are nonzero.

A trivial solution to Eq. \eqref{constant} is the $\numu$-$\nutau$ symmetry, for which we have
\bea
\alpha = 0, ~~~~ \beta= 1, ~~~~ \gamma= -1\,.
\eea
With this $\nu_\mu$-$\nu_\tau$ symmetry, one will always get the same flavor ratios at Earth for $\nu_\mu$ and $\nu_\tau$, 
regardless of the flavor ratios at cosmic sources; 
the Earthly ``triangle'' collapses to the vertical line bisecting the $w_\mu$ and $w_\tau$ coordinates.

\subsubsection{Broken $\nu_\mu$-$\nu_\tau$ Symmetry and Vanishing Det({\boldmath{$\scrP$}})}
\label{subsubsec:NoMuTauSym}
%
With the help of unitary conditions and linear algebra simplifications, we find
%
\beq{DetSimplified}
{\rm Det}\left(|U|^2\right) = \left|
\ba{ccc}
1 & 1 &\ \ \  3\ \ \\
 & & \\
|U_{\mu 1}|^2-|U_{\tau 1}|^2 &\ \  |U_{\mu 2}|^2-|U_{\tau 2}|^2 &\ \ \  0\ \ \\
 & & \\
|U_{\tau 1}|^2 & |U_{\tau 2}|^2 &\ \ \  1\ \ \\
\ea
\right|\,.
\eeq
%
From this simplified expression, we can readily obtain
\bea
\label{DetSimplified2}
{\rm Det}\left(|U|^2\right) = \cos2\theta_{13}\cos2\theta_{21}\cos2\theta_{32} -\frac 1 2 \sin2\theta_{21}\sin2\theta_{32}(1-3\sin^2 {\theta_{13}})\sin\theta_{13}\cos \delta.
\eea
As a check, we see that this expression vanishes under exact $\nu_\mu$-$\nu_\tau$ symmetry,
which requires $\sin 2\theta_{21}\sin \theta_{13}\cos \delta=0$ and $\theta_{32}=\frac \pi 4$. 
What we now point out that is new, is that a vanishing determinant is also viable even when the $\nu_\mu$-$\nu_\tau$~symmetry is broken.
Setting the determinant in Eq.~\rf{DetSimplified2} to zero, we solve for
the CP-violating phase $\delta$ as determined by the three mixing angles.  
We find that Det$(|U|^2)$ vanishes, and therefore Det$(\scrP)$ vanishes, if  
\bea
\label{deltaTuned}
\cos{\delta}=2\cot2\theta_{32}\cot2\theta_{21}\cot2\theta_{13}\,\left( \frac {2\cos{\theta_{13}}} {1-3\sin ^2 {\theta_{13}}}\right)\,.
\eea
With the recent global best fit data~\cite{Forero:2014bxa,Gonzalez-Garcia:2014bfa,NuFit} one can easily find 
that the RHS of Eq.~\eqref{deltaTuned} ranges from -0.92 to 0.91 as the mixing angles vary in the $2\sigma$ range, 
which corresponds to the ranges for $\delta$ of $\pm[24.5^{\circ}$, $156.9^{\circ}]$.
Currently, the Dirac CP-violating phase is still largely unconstrained at a $2\sigma$ level. 
Should terrestrial experiments in the future infer a $\delta$ satisfying Eq.~\eqref{deltaTuned}, 
then Det($\scrP$) vanishes and the inverse propagation matrix $\scrP^{-1}$
needed to evolve the observed neutrino flavor ratio backwards to their injection ratios at cosmic sources
does not exist. 

\begin{figure}[t!]
\includegraphics[height=4.0cm, width=4.5cm]{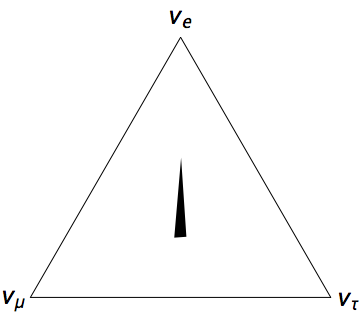}
~~~~~~
\includegraphics[height=4.0cm, width=4.5cm]{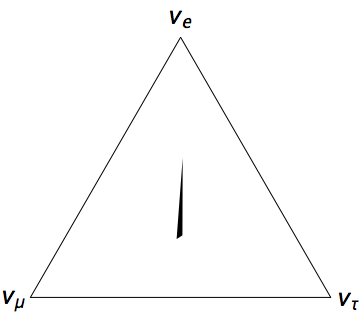}
~~~~~~
\includegraphics[height=4.0cm, width=4.5cm]{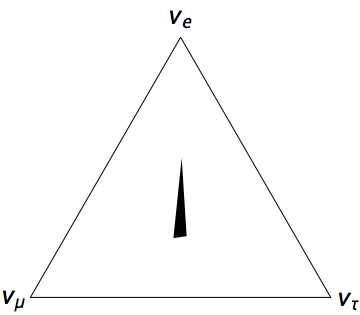}
~~~~~~
\caption{The Earthly triangles for the best values of, from left to right, 
the Normal Hierarchy with $\theta_{32}$ in first octant,
Normal Hierarchy with $\theta_{32}$ in second octet, 
and the Inverted Hierarchy.}
\label{fig:best-fits}
\end{figure}

\subsubsection{A Theorem Relating the Area of the Earthly Triangle and Det(\boldmath{$\scrP$}) }
\label{subsubsec:theorem}

In this section, we present and prove an interesting theorem that relates the determinant of the $\scrP$~matrix 
to the area of the allowed flavor triangle on Earth. 
In 3D space, the allowed area of the original flavor triangle is $\frac{\sqrt{3}}{2}$.  
The equilateral hyper-triangle results from the single unitary constraint, $W_e+W_\mu+W_\tau=1$,
on infinite 3D Euclidean space.
After mixing, the area $S$ of the Earthly flavor ratio triangle with the three vertices given by 
$\scrP\,(1,0,0)^{\rm T}$, $\scrP\,(0,1,0)^{\rm T}$, and $\scrP\,(0,0,1)^{\rm T}$,
is much smaller than the area ($\frac{\sqrt{3}}{2}$) of the original source triangle.
Interestingly, we find that the area of this Earthly triangle, denoted by $S$, 
is proportional to the absolute value of the determinant of the $\scrP$ matrix. 
Simply stated, the theorem says that 
\beq{theorem}
{\rm THEOREM:}\quad\quad\quad
 S=\frac{\sqrt{3}}{2} \, | {\rm Det}(\scrP) | \,.
\eeq
In terms of the three independent off-diagonal elements of $\scrP$ defined in Eq.~\rf{P-sym}, 
we have 
\beq{Det-P}
{\rm Det}(\scrP)=1-2(a+b+c)+3(ab+ac+bc).
\eeq
Also in terms of these three independent parameters,
the coordinates of the three vertices of the Earthly triangle are ($1-a-b$, $a$, $b$), ($a$, $1-a-c$, $c$), and ($b$, $c$, $1-b-c$). 
Taking the difference of these three vectors defines the vector lengths of the three sides of the Earthly triangle.
And from the lengths of the three sides, we calculate the triangle's area, $S$.
A bit of algebra leads to $\sqrt{3}/2$ times the expression in Eq.~\rf{Det-P}.
The theorem is proven.

An interesting case is the area of the quark-flavor triangle after mixing.
Inputing the quark sector values $a=9.65\%$, $b=0.017\%$, and $c=0.33\%$, 
we get for the area $A_{\rm mixed} = 80.1\%$
times the unmixed (original) area of $\sqrt{3}/2$.
Very little contraction of the triangle's area has occurred, due to very little mixing.
The situation is very different with the large mixing angles in the neutrino sector,
as we shall see.

Illuminating checks result for the TBM model, where the $\mutau$~symmetry (which implies a vanishing determinant)
leads to an Earthly flavor ratio located on the $w_\mu=w_\tau$ symmetry line; 
and for the even simpler case of the democratic propagation matrix, $\scrP_\third$, 
where the two constraints  of the exact $\nu_e$-$\mutau$~symmetry 
(again implying a vanishing determinant) lead to a single point at the centroid of the injection triangle. 
In both of the above cases, the area of the region occupied by the Earthly triangle is zero.

\subsection{Extra Constraint -- Second Example: Thinness of the Earthly Triangle}
\label{subsec:thinness}
\begin{figure}[t!]
\includegraphics[height=6.5cm, width=7.5cm]{NH1_BFtriangle.png}
~~
\includegraphics[height=6.5cm, width=7.5cm]{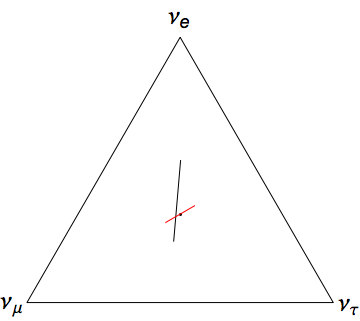}
~~~~~~
\caption{The left panel shows the flavor triangle at Earth, for the best fit values of the Normal Hierarchy with $\theta_{32}$ in the first octant 
(also shown in the previous Fig.~\ref{fig:best-fits}),
and the right panel shows the straight line that results from the assumption that  $\nu_\tau$ production at the source is negligible.
Note that the ``no-$\nutau$ injected'' line is a boundary of the full triangle, 
and therefore does not contain the centroid (shown) at the point ($\third,\ \third,\ \third$).
The small red line is the 2$\sigma$ statistical error at the centroid values, assuming 140 total events. 
We note that the statistical error bar is comparable to the width of the Earthly triangle, shown in the left panel.}
\label{fig:no-tau-line}
\end{figure}

The difference between the vertices of the Earthly triangle which are closest to each other, i.e. the points 
$\scrP (0,\ 1,\ 0)^{\rm T}$ and $\scrP (0,\ 0,\ 1)^{\rm T}$, projected onto the $w_e=0$ axis of the triangle, 
is
\beq{trianglewidth}
\Delta W(w_e=0) = 2-(a+b+4c) \,.
\eeq
%
This final expression turns out to be 2/9 times the parameter ``$C$'' of Ref.~\cite{Pakvasa:2008zz}, 
where it was shown that $C$ is second order in deviation of the standard 
angle-phase parameters $\theta_{13}\cos\delta$ and $\theta_{32}$ 
from their respective TBM values of zero and $\frac{\pi}{4}$, and independent of $\theta_{21}$.
Thus we learn that the small deviations from TBM will appear as an even smaller deviation of the Earthly triangle from the 
straight line of the TBM model.
Furthermore, the long sides of the Earthly triangle will in general be slightly rotated with respect to the vertical TBM line 
(by small angles that are second order in $\theta_{13}$ and ($\theta_{32}-\frac{\pi}{4}$)).
Examples of the thinness of the Earthly triangle, and its rotation, are shown in Fig.~\ref{fig:best-fits} 
for the best fit values of the Normal Hierarchy with $\theta_{32}$ lying in the first and second octants, 
and for the Inverted Hierarchy. 
The example of the best-fit values for the Normal Hierarchy with $\theta_{32}$ in the first octant is shown again in Fig.~\ref{fig:no-tau-line}.
In particular, the fraction of original triangular area given by the Earthly triangle ranges from a maximum of 1.2\% at $\delta=0$, 
to zero for the vanishing determinant value discussed above.
The ability of an experiment to measure the width of the Earthly triangle thus requires a very good accuracy, 
of order~$(\Delta W(w_e=0) )$, as given above.

%

\subsection{Extra Constraint -- Third Example: \boldmath$\nutau$ Confusion, and Statistical Error}
\label{subsec:2wandStatErrors} 
\subsubsection{Tau Neutrino Interaction Topologies, and Ambiguity}
\label{subsubsec:nutau}
 The mean free decay length of the tau in vacuum (\ie $c\times$lifetime) is 
$L_\tau = \quarter\,(E_\tau/5\,{\rm PeV})$~km, and so some tau track lengths may be visible (leading to so-called ``double-bang'' events)
in the energy decade centered on $\sim 5$~PeV in IceCube, and at somewhat higher energy in the 
proposed larger, sparser, IceCube-Gen2~\cite{Aartsen:2014njl}.
We pose our discussion on the IceCube configuration of optical modules.
Below about 1~PeV, the tau decays promptly, and so its decay shower contributes the initial $\nutau$ shower.
The event then adds to the ``shower'' class of events, like the $\nue$.
Well above about 10~PeV, the $\nutau$ shower and the tau decay shower are well separated, so much so that 
both showers do not appear in the detector.
If the $\nutau$ shower appears in the detector (with $\sim$ 20\% of the initial neutrino energy), then it is followed by a track,
adding to the ``track'' event class, like the charged-current $\numu$ events.
If the tau shower appears in the detector, it is preceded by a track, in which case the event is identifiable as a 
$\nutau$-initiated event.  The isolation of event types depends very much on the particular detector's
configuration, its efficiency for separation of events in classes, and the energy of the initial neutrino.
Accordingly, the pigeon-holing of event types is bet left to the experimenters.
Here we analyze just the simple case of lower energy events, $\lsim$~PeV, where the the showering fraction $\wsh$ is identified with 
the sum $w_e + w_\tau$, and the track fraction $\wtr$ is identified with the $w_\mu$.  
(We neglect neutral currents, since they contribute a lower-energy shower, 
which is negligible for a sufficiently falling energy-spectrum of initial neutrinos.)
We remind the reader that the highest energy of neutrino events measured to date is only 2~PeV.

For our model calculation, the separation into $\nue$ and $\nutau$ events cannot be made.
There are just two track topologies to consider, $\wtr=w_\mu$ and $\wsh=w_e+w_\tau$,
 with $\wtr+\wsh=1$.
There is a single parameter to be inferred from experiment, $w_\mu$, or equivalently, $\wsh=1-w_\mu$.

\subsubsection{Statistical Error}
\label{subsubsec:stats}
One may ask what kind of statistical errors are expected from the measurement of a finite event number at Earth.
To simplify the discussion we will assume that $\nue$ and $\nutau$ contribute only showering events, and $\numu$ contributes 
only track events, i.e., we neglect the small contribution to shower events from $\numu$ neutral-current interactions,
and the contribution to track events from $\nutau$ events. 

\begin{figure}[t!]
\includegraphics[height=6.5cm, width=10.5cm]{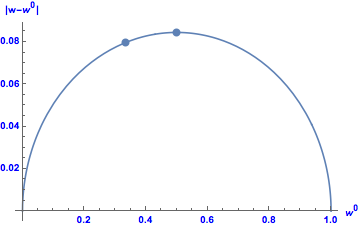}
\caption
{The $2\sigma$ ($m$=2, 95\% CL) standard deviation from Eq.~\rf{stat4} for $\Ntot=140$ events,
versus the measured $w^0$ (the same for $\wtr$ and $\wsh$).
The dots denote the special values of $w^0 = 1/2$ where the standard deviation has a maximum, 
and $w^0 = 1/3$, the centroid value expected for neutrino flavors from the pion decay chain.}
\label{fig:errorfactor}
\end{figure}

The statistical error will depend on the total number of events, $\Ntot$, and on the measured number of
track events (note that the measured number of shower events is related by 
$\Ntot = \Ntr+\Nsh$, where the experiment partitions the total number of events into ``track'' and ``shower'' events.)
We begin with the definition $\wtr^0 =\Ntr/\Ntot$, where the superscript ``0'' on $w$ denotes the measured ratio. 
Then 
\bea
\label{stats1}
\ln(\wtr) = \ln(\Ntr)-\ln(\Ntr+\Nsh)\,, \quad {\rm and\ therefore} \nonumber   \\ 
\nonumber \\
\delta \ln \wtr^0 = \frac{\delta\wtr}{\wtr^0}=\frac{1}{\Ntot}\left( \delta\Ntr \left(\frac{\Nsh}{\Ntr}\right)-\delta\Nsh \right)\,.
\eea
The two uncertainties, $\delta\Ntr$ and $\delta\Nsh$, are statistically uncorrelated, so we add them in quadrature.
We also invoke the Poisson result that $\delta\Ntr=\sqrt{\Ntr}$ and $\delta\Nsh=\sqrt{\Nsh}$.
Then a bit of algebra returns the relative error
\beq{stat15}
\left| \frac{\delta\wtr}{\wtr^0} \right| = \sqrt{\frac{\Nsh}{\Ntr\,\Ntot} } \,,
\eeq
and the absolute error
\beq{stat2}
|\delta\wtr| = \sqrt{\frac{\Ntr\,\Nsh}{\Ntot^3} } = \sqrt{\frac{\wtr^0\,(1-\wtr^0)}{\Ntot} } \,.
\eeq
 The $\Ntr$-$\Nsh$ symmetry displayed in the middle result of Eq.~\rf{stat2} tells us that 
 the errors in $|\delta\wtr |$ and in $|\delta\wsh |$ are the same 
 (\ie that the two-dimensional error ellipse is in fact a circle.).  
 Thus we may use the one d.o.f. formula (not surprising, since we have the constraint $\wtr+\wsh=1$)
 for the $m\sigma$ error contour:
 \beq{stat3}
 \frac{(\wtr-\wtr^0)^2}{\delta\wtr^2}=m^2\,.
 \eeq
Finally we arrive at the final formula for the statistical error:
\beq{stat4}
| \wtr-\wtr^0 | = m\;\sqrt {\frac{\wtr^0\,(1-\wtr^0)}{\Ntot} }\,.
\eeq
The factor $\wtr^0\,(1-\wtr^0)$ has a maximum of 1/4 at its symmetry point, $\wtr^0=1/2$.
In Fig.~\ref{fig:errorfactor} we plot this factor as a function of $\wtr^0$.
In the figure, we have marked the symmetry point at $\wtr=1/2$,
and also the democratic, centroid value $\wtr=1/3$.

In the right panel of Fig.~\ref{fig:no-tau-line} we show with a small red line, 
the resulting $2\sigma$ error ($m=2$, $95\%$~CL) for the democratic value $\wtr=1/3$.
We show the error to lie along the bisector of the $w_e$-$w_\tau$ axis;
in fact, the error is the same anywhere along the oblique line of constant $w_\mu$.
For the plot, we have taken $\Ntot=140$~events, the number that would typically 
be collected by IceCube in ten years ($\sim 14$~events per year), 
or the extension Gen2 in two years ($\sim 70$~events per year).
The value of the error is small, $\pm 2\,\sqrt{ 1/3\times 2/3 \div 140}=\pm 0.080$. However, as seen in the figure, this statistical error is not small enough to allow differentiation of the 
Earthly flavor triangle's centroid and $W_\tau=0$~boundary,
nor to allow a clean inference of the Earthly triangle's width, at $95\%$~CL.
Of course, ten years of data collection by Gen2 will reduce the two-year error calculated here 
with a factor of $1/\sqrt{5} = 0.45$ or more.

The error for $\wsh$ is the same as that for $\wtr$, 
but cannot be identified with a unique point in the flavor triangle since we have assumed here that 
$w_e$ and $w_\tau$ are measured only in their summation: $w_e + w_\tau=\wsh^0 = 1-\wtr^0$.

\subsection{Extra Constraint -- Fourth Example: No \boldmath$\nutau$'s Produced at the Cosmic Source}
\label{subsec:nonutau}
Conventional dynamics at the neutrino sources will not produce many $\nutau$'s,
because of the large mass of the leptonic partner particle, the $\tau$ ($m_\tau=1.777$~GeV).
$D_s$ production and decay will result in a few $\nutau$'s, with 
the expectation of 0.1\% for $W_\tau$~\cite{pakvasa:2007dc}. Setting $W_\tau$ to zero is a constraint, reducing the Earthly triangle to a straight line joining the points at 
$\scrP\,(1,0,0)^{\rm T}$ and $\scrP\,(0,1,0)^{\rm T}$.
This line is an edge of the Earthly triangle obtained with arbitrary injection models.
As such, the line should not contain the Earthly triangle's centroid.

An interesting application of our ``area theorem'' is that at least one of the following popular arguments must be invalid: \\
(i)  no $\nu_\tau$'s are produced at the cosmic source; \\
(ii) ${\rm Det}(\scrP) \neq 0$; \\
(iii) the flavor ratio at Earth is democratically $\third$(1:1:1). \\
We note that the IceCube experiment claims compatibility of its data with the roughly democratic prediction from the pion decay-chain.

We construct the proof by showing that (iii) is invalid if (i) and (ii) are correct. 
Once this is proven, we are done: not all three statements may be correct.

According to Eq.~\rf{theorem},  
a non-vanishing determinant of the $\scrP$ matrix ( (ii) above) 
guarantees that all the possible Earthly points occupy a small triangle with non-vanishing area. 
Moreover, with negligible $\nu_\tau$ produced at cosmic sources~\cite{Waxman:1997ti} ( (i) above), 
the small triangle is further reduced, to a straight line which connects the two points 
propagated from the vectors $\vec{W}=(1,0,0)$ and $\vec{W}=(0,1,0)$.
Recall that the point $\third$(1:1:1) is the centroid of the small triangle. 
The centroid can never be a point on the boundary of the Earthly triangle, as it is an interior point. 
Thus it cannot be the Earthly point resulting from omission of $\nutau$ injection.
Thus, our third argument (iii) is invalid. 

Working backwards through the logic then leads to the inference that 
any observation of a strict democratic ratio on Earth 
would implicate either a vanishing determinant of $\scrP$, 
or some nonzero $\nu_\tau$ production in the injection model. 
And to the inference that 
if the $\scrP$ matrix has a nonzero determinant, then
any injection model with one or more flavors vanishing can never lead to a democratic ratio on Earth, 
since  then the propagated point arriving at Earth 
must be located on the triangle's edge (boundary),
which does not include the triangle's centroid.

\subsubsection{Some Implications of Potential Flavor Measurements}
\label{subsubsec:FlavorMeasure}
As terrestrial experiments are making more precise measurements of the four leptonic mixing parameters, 
we can expect the accuracy of the $\scrP$ matrix to continually increase.
Hence we may well determine the propagated Earthly triangle, and ``no-$\nu_\tau$ injected'' boundary line, in the future~\cite{Mena:2014sja,*Palomares-Ruiz:2015mka}. 
On the observational side, although the number of events expected in the near term by neutrino telescopes will not allow 
flavor analyses to determine the position of the flavor-ratio point at Earth 
(see, e.g. the statistical error bar presented in Fig.~\ref{fig:no-tau-line}), the point is determinable in the long term.
Assuming statistically large future event samples, we infer an interesting and distinct result of each of the three possible locations of the flavor point with respect to the triangle: 
\begin{itemize}  
\item{The measured Earthly point lies inside the Earthly triangle}. \\
In this case, the point is not on a boundary, and in particular is not on the ``no-$\tau$ injected'' boundary.
So a significant amount of $\nutau$ must be emitted at the sources.
\item{The measured Earthly point lies on the ``no-$\nutau$ injected'' boundary}. \\
Then it is necessarily so that the sources do not emit a significant amount of $\nutau$.
\item{The measured Earthly point lies outside the Earthly triangle}. \\
In this case, the implication is that some exotic physics must come into play.
Examples of exotic physics include neutrino decay~\cite{Pakvasa:2012db,*Beacom:2002vi,*Berryman:2014qha}, 
active-sterile neutrino mixing~\cite{Berryman:2014yoa}, 
and new neutrino interactions~\cite{Ioka:2014kca,*Ohlsson:2013vaa,*Mocioiu:2014gua,*Ng:2014pca}.
\end{itemize}

In fact the point in the Earthly flavor triangle characterizing flavor ratios is likely to be energy-dependent.
It is beyond the scope of this paper to deal with the extra complications that may result.
When more flavor data is available, then it may be warranted to include these complicating options.
For now the event sample is sufficient for only primitive flavor analyses.

\subsubsection{Small but Nonzero $W_\tau$}
\label{subsubsec:smalltau}
\begin{figure}
\centering
\includegraphics[height=6.5cm, width=7.5cm]{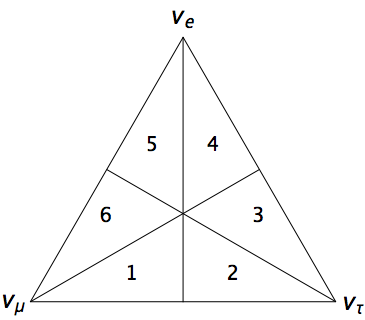}
~~~~~~
\caption{Source triangle partitioned into six sub-triangles, each characterized by a distinct ordering of the injection ratios 
$\{W_e,\ W_\mu,\ W_\tau\}$, as well as the Earthly ratios $\{w_e,\ w_\mu,\  w_\tau\}$.}
\label{fig:six}
\end{figure}

%
There are a priori six possible orderings of the three injection flavor ratios $\{W_e,\ W_\mu,\ W_\tau\}$ 
and of the three Earthly flavor ratios $\{w_e,\ w_\mu,\ w_\tau\}$.
When the cosmic triangle is divided by the three bisectors, as shown  in Fig.~\ref{fig:six},
there results six symmetric sub-triangles. The six sub-triangles meet at the geometric centroid ($w_e = w_\mu = w_\tau = \third$).
Each bisector divides the ordering of two of the three $W_\alpha$'s or $w_\alpha$'s.
Thus, there is a 1-1 correspondence between the orderings of the $W_\alpha$'s and $w_\alpha$'s,
and the regions of the six sub-triangles.
The correspondence is: \\
Sub-triangle 1:  $W_\mu \ge W_\tau \ge W_e$ and $w_\mu \ge w_\tau \ge w_e$; \\
Sub-triangle 2:  $W_\tau \ge W_\mu \ge W_e$ and $w_\tau \ge w_\mu \ge w_e$; \\
Sub-triangle 3:  $W_\tau \ge W_e \ge W_\mu$ and $w_\tau \ge w_e \ge w_\mu$; \\
Sub-triangle 4:  $W_e \ge W_\tau \ge W_\mu$ and $w_e \ge w_\tau \ge w_\mu$; \\
Sub-triangle 5:  $W_e \ge W_\mu \ge W_\tau$ and $w_e \ge w_\mu \ge w_\tau$; \\
Sub-triangle 6:  $W_\mu \ge W_e \ge W_\tau$ and $w_\mu \ge w_e \ge w_\tau$.\\
When ${\rm Det}(\scrP) \neq 0$, the Earthly triangle must have representation in each of the six sub-triangular regions, 
since the triangle must have a nonzero area and it must contain the centroid.

Popular injection models describe sources or processes that produce very little $\nutau$ i.e., $W_\tau < \{W_e, W_\mu\}$. Thus, there are only two popular orderings for the injection ratios:
(i) $W_e> W_\mu> W_\tau$, and (ii) $W_\mu> W_e> W_\tau$. 
The first ordering is that of sub-triangle number 5;
the second ordering is that of sub-triangle number 6.

We may ask whether any of these orderings would be preserved when the neutrinos are propagated to Earth. 
The answer is yes, the ordering $W_e > W_\mu$ (i.e., $W_\mu/W_e < 1$) is preserved at Earth, 
although the ordering $W_\mu>W_e$ (i.e., $W_\mu/W_e > 1$) need not be.
Here is the proof:\\
from Eq.~\rf{P-sym}, we get  
\beq{wdiff}
w_e - w_\mu = x\,(W_e-W_\tau)-y\,(W_\mu-W_\tau)\,,
\eeq
where $x\equiv (1-2a-b)$, and $y\equiv (1-2a-c)$.
Earlier we stated that $c > a{\rm\ or\ }b$, and that $1 > a+b+c$, even when the $\scrP$~matrix is perturbed about its TBM value.
Thus, $x > y > 0$.
Now assume that $W_e > W_\mu> W_\tau$.
Then $(W_e-W_\tau) > (W_\mu-W_\tau) > 0$.
Hence, the RHS of Eq.~\rf{wdiff} is positive, and so the LHS must be positive, i.e., 
$w_e > w_\mu$, and the flavor ordering is maintained. Next assume that $W_\mu> W_e > W_\tau$.
Then the RHS is of indefinite sign, and therefore so is the LHS. Contrapositive argument infers that any Earthly point found inside sub-triangle 1, 2, or 6, must be due 
to an injection model with $W_\mu > W_e > W_\tau$ (corresponding to sub-triangle 6).

In fact, the $W_\mu/W_e$ ratio is readily measured at a neutrino telescope.
In~\cite{Fu:2012zr}, a relation was derived for this ratio.  
Here we find that the relation may be simplified in form to
\beq{emu_ratio}
\frac{W_\mu}{W_e} = \frac {\scrP_{e\mu}-w_\mu} {w_\mu-\scrP_{\mu\mu}} \,. 
\eeq
The parameters $\scrP_{e\mu}$ and $\scrP_{\mu\mu}$ are determined from terrestrial measurements of mixing angles, 
as should be evident from this paper.  So just the parameter $w_\mu$ remains to be inferred from experiment.  
For neutrino energies $\lsim$~PeV, the only track events are produced by the $\numu$ charged current,
and so Eq.~\rf{emu_ratio} is readily determined.  
At a neutrino telescope such as IceCube, the fraction of $\numu$ events incident at Earth $w_\mu$ 
is (neglecting the neutral-current contribution) also the fraction of track events.
We emphasize that this result, and the derivation of it, remain valid independent of whether the 
determinant of $\scrP$ is vanishing or non-vanishing.

%

\section{Conclusions}
\label{sec:conclusions}
Flavor evolution of active neutrinos from distant astrophysical sources depends on the three mixing angles and one CP-violating phase, 
but in just three and not four independent CP-conserving combinations. 
This is because the large distance effectively averages over oscillation phases,
reducing the quantum mechanical probability for flavor oscillation to a simpler classical mixing probability.
This reduction entitles us to conveniently encapsulate the evolution in a symmetric 3~by~3 ``flavor propagation matrix'' 
$\scrP\equiv \langle P\rangle_{\rm phase\ averaging}$. 
Unitarity of the PMNS leptonic mixing matrix implies certain restrictions on the $\scrP$~matrix.
We have incorporated and explained these restrictions.

If the $\scrP$~matrix has nonzero determinant, then it may be inverted and the neutrino flavor ratios at cosmic sources
may be directly inferred from flavor ratios observed at Earth.  A good approximation to the $\scrP$~matrix is 
the $\mutau$~symmetric TBM matrix  $\scrP_{\rm TBM}$ with vanishing determinant.  
However, the form of the TBM matrix is known to be not strictly that of Nature, and so the question arises,
``with $\mutau$~symmetry broken, can we conclude that Nature's $\scrP$~matrix has nonzero determinant and so is invertible?''
We showed that the answer is negative. There is a unique value of $\delta$, 
given in terms of the three angle parameters $\{\theta_{32},\ \theta_{12},\ \theta_{13}\}$,
for which Det($\scrP$) is vanishing.  
If further experiments were to establish this value of $\delta$ as correct, 
then some new symmetry should be sought to enforce ${\rm Det}(\scrP)=0$ for the neutrino sector.

Next we proved a theorem, that the determinant of $\scrP$ is proportional to the area of the Earthly flavor triangle.
Thus, the inversion of $\scrP$ depends on the Earthly triangle having a nonzero area.
The TBM version of $\scrP$ has a vanishing determinant and consequently, a vanishing area for its Earthly triangle.
Thus, the small deviations of Nature's choices from the TBM values lead to a small nonzero area.
We quantified this statement.

We then considered a simple model which allowed a straightforward calculation of statistical significance.
The model was that $\nutau$ events contributed only to shower topologies in the detector,
as would be expected for an event sample with energy up to $\sim$PeV.  
Thus, the origin of the shower events is the sum of $\nutau$ and $\nue$ events, while the origin of the track events is purely $\numu$.
This model does not allow $\nue$--$\nutau$ event separation, and so even idealized Earthly measurements within this paradigm 
cannot determine a point in the Earthly flavor triangle, but only a point on the line parameterized by $\wsh$ and $\wtr=1-\wsh$.
We analyzed the dependence of the experimentally-determined point on the line on the event statistics.
We established $2\sigma$, 95\% CL errors for an assumed 140 total events, a number expected to be 
typical of a decade of IceCube measurements, or a year or two of the IceCube extension Gen2.
We found that the statistical error in this case to be comparable to the width of the Earthly triangle.
Thus, separation of $w_e$ from $w_\tau$ requires improved statistics.
Above about a~PeV, the topology of the $\nutau$ events becomes complicated, 
due to the separation lengths between the resulting ``double bangs'' compared to the 
experiment's photo-detector separation length(s).  The former lengths are stochastically distributed,
and the mean separation length is energy-dependent.
A three-flavor analysis under these conditions is beyond the scope of this paper, and, in fact, best left to the experimenters.

The assumption that $\nutau$'s are not produced at the sources, \ie that $W_\tau=0$, 
is commonly believed to be true, due to the heavy mass of the tau particle associated with the $\nutau$ in charged--current production.  
Accordingly, we next considered the implications of the lack of significant $\nutau$ production at the source, by setting $W_\tau$ to zero.  
The condition $W_\tau=0$ reduces the injection flavor triangle 
and the Earthly flavor triangle to straight lines on the boundaries of the would-be triangles.  
Whether the observed flavor point lies on the inside of the ``no--$\nutau$'' boundary line, outside the line, or on the line
leads to distinct physics conclusions, as we described. 
Even were $\nutau$ injection to be significant but less than that of $\numu$ and $\nue$, 
we showed that with Earthly flavor measurements, a test of source orderings 
$W_\mu > W_e > W_\tau$ versus $ W_e  > W_\mu >  W_\tau$ becomes possible.
 
We conclude that although the Earthly flavor triangle is greatly reduced in area from the injection flavor triangle,
it offers a powerful tool to elucidate at Earth some details of astrophysical neutrino injection. 

\section*{Acknowledgments}

We thank W. Rodejohann for pointing out an error in our first arXiv'd version of this paper and acknowledge useful discussion with S. Palomares-Ruiz on the nature of track-to-shower ratios
at IceCube.
LF and TJW are supported in part by US DoE grant DE-SC0011981, and in part by NSF grant PHY11-25915  
the Kavli Institute for Theoretical Physics, Santa Barbara. 
TJW is also supported by a Simons Foundation Grant, \#306329, and acknowledges support from the Munich Institute for 
Astro- and Particle Physics (MIAPP) of the DFG cluster of excellence ``Origin and Structure of the Universe"
during the preparation of this work.

\bibliography{reference}
  
\end{document}